\documentclass[twocolumn,showpacs,aps,prl,nofootinbib,superscriptaddress]{revtex4}

\usepackage{xspace}
\usepackage{graphicx}
\usepackage{dcolumn}
\usepackage{amsmath}
\usepackage{epsfig}

\input babarsym.tex
\def\BRtaumtoKmpiz           {\ensuremath{\BR(\taumtoKmpiz)}\xspace}

\def\Ntautau                 {\ensuremath{N_{\scriptscriptstyle{\tau\tau}}}\xspace}

\def\SigEff                  {\ensuremath{\varepsilon_{\scriptscriptstyle{\rm sig}}}\xspace}

\def\rhom                    {\ensuremath{\rho^-}\xspace}

\def\taumtoKmpiz             {\ensuremath{\taum \to \Km \piz \nut}\xspace}

\def\strangemass             {\ensuremath{m_{\scriptscriptstyle \s}}\xspace}

\def\Kstar                   {\ensuremath{K^{*}}\xspace}
\def\Kstarp                  {\ensuremath{K^{*+}}\xspace}

\def\Kstareightninetwo       {\ensuremath{K^{*}(892)}\xspace}

\def\Kstarfourteentenm       {\ensuremath{K^{*}(1410)^{-}}\xspace}

\def\Kstarfourteenthirtym    {\ensuremath{K_{0}^{*}(1430)^{-}}\xspace}

\def\dz                      {\ensuremath{d_{\scriptscriptstyle 0}\xspace}}

\def\aleph                   {ALEPH}
\def\cleo                    {CLEO}
\def\opal                    {OPAL}

\def\kk                      {{\tt KK2f}}
\def\tauola                  {{\tt TAUOLA}}

\def\ckm                     {CKM}

\def\etag                    {$e$-tag}
\def\mutag                   {$\mu$-tag}

\def\taubkgds                {\mtau\ backgrounds}

\def\mc                      {Monte Carlo}

\def\Cerenkov                {Cherenkov}

\def\crosssection            {cross-section}

\newcommand{\BABARPubYear}     {07}
\newcommand{\BABARPubNumber}   {036}

\newcommand{\SLACPubNumber} {12681}

\begin{document}

\preprint{\babar-PUB-\BABARPubYear/\BABARPubNumber}
\preprint{SLAC-PUB-\SLACPubNumber}

\begin{flushleft}
\babar-PUB-\BABARPubYear/\BABARPubNumber \\
SLAC-PUB-\SLACPubNumber
\end{flushleft}

\title{
{\large \bf \boldmath
Measurement of the \taumtoKmpiz\ Branching Fraction 
}
}

\date{\today}

%% author list as of 06-Jun-2007 (572 authors)
%
\author{B.~Aubert}
\author{M.~Bona}
\author{D.~Boutigny}
\author{Y.~Karyotakis}
\author{J.~P.~Lees}
\author{V.~Poireau}
\author{X.~Prudent}
\author{V.~Tisserand}
\author{A.~Zghiche}
\affiliation{Laboratoire de Physique des Particules, IN2P3/CNRS et Universit\'e de Savoie, F-74941 Annecy-Le-Vieux, France }
\author{J.~Garra~Tico}
\author{E.~Grauges}
\affiliation{Universitat de Barcelona, Facultat de Fisica, Departament ECM, E-08028 Barcelona, Spain }
% ...
\collaboration{The \babar\ Collaboration}
\noaffiliation

\begin{abstract} 
A measurement of the {\ensuremath{\tau^- \rightarrow K^- \pi^0 \nu_{\tau}}\xspace}
branching fraction has been made using $230.2 {\ensuremath{\mbox{\,fb}^{-1}}\xspace}$ of
data recorded by the {\mbox{\slshape B\kern-0.1em{\smaller A}\kern-0.1em B\kern-0.1em{\smaller A\kern-0.2em R}}}
detector at the {PEP-II} {\ensuremath{e^+e^-}\xspace} collider, located at the
Stanford Linear Accelerator Center (SLAC), at a center of mass energy $\sqrt{s}$ close to $10.58 {\ensuremath{\mathrm{\,Ge\kern -0.1em V}}\xspace}$. 
We measure ${{\ensuremath{\cal B}\xspace}}({\ensuremath{\tau^- \rightarrow K^- \pi^0 \nu_{\tau}}\xspace}) = 
(0.416 \pm 0.003 \, {\ensuremath{\mathrm{(stat)}}\xspace} \pm 0.018 \, {\ensuremath{\mathrm{(syst)}}\xspace})\%$.
\end{abstract}

\pacs{13.35.Dx, 14.60.Fg, 11.30.Hv}

\maketitle

The \mtau\ is the only lepton with a sufficiently large mass to decay to hadrons.
Tau decays to hadronic final states proceed via $W$ exchange and thus 
the decay rates to the final states containing a strange quark is suppressed by the factor 
\ensuremath{\left( \Vus / \Vud \right)^{2}} relative to the non-strange final states, where 
$\Vud$ and $\Vus$ are the moduli of the Cabibbo-Kobayashi-Maskawa (\ckm)
matrix~\cite{Cabibbo:1963yz, Kobayashi:1973fv} elements. 
For a given value of \strangemass~\cite{Jamin:2006tj}, \Vus can be determined 
up to unprecendented precision~\cite{Gamiz:2002nu,Maltman:2007pr}
from the inclusive sum of the branching fractions of $\tau$ decays to
hadronic final states with net strangeness equal to unity. 
This determination of \Vus can be perfomed even without
detailed knowledge of the hadronic mass spectrum arising
due to incompleteness in our understanding of some of the
intermediate resonance contributions.

In this paper we present a measurement of the \taumtoKmpiz\ branching 
fraction\footnote{Throughout this paper, the charge conjugate decays are implied.}.
In recent years, measurements of the branching fractions for \mtau\ decays to strange 
hadronic final states have been made using \cleo\ \cite{Battle:1994by}, \aleph\ \cite{Barate:1999hi}
and \opal\ \cite{Abbiendi:2004xa} data, but have often been limited by the size of the available data samples. 
The high luminosity provided by the \pep2\ asymmetric-energy $e^+ e^-$ storage 
rings \cite{:1993mp} at the Stanford Linear Accelerator Center (SLAC) coupled with the large \crosssection\ 
for $\tau^+ \tau^-$-pair production has given us a very large sample for studying such decays in 
the \babar\ detector. 

The \babar\ detector is described in detail elsewhere~\cite{Aubert:2001tu}. 
Charged particles are detected and their momenta measured with a five-layer 
double-sided silicon vertex tracker (SVT) and a 40-layer drift chamber (DCH)
inside a $1.5 \rm{\, T}$ superconducting solenoidal magnet. A ring-imaging 
\Cerenkov\ detector (DIRC) provides additional separation power for identification 
of charged particles for momenta greater than $1~\gevc$, and thus complements \dedx\ 
measurements in the DCH, useful for the identification of charged particles below $1~\gevc$. 
Energies of photons and electrons are measured by a CsI(Tl) crystal electromagnetic calorimeter 
(EMC), and the instrumented magnetic flux return (IFR) is used to identify muons.

The analysis described in this paper is based on a data sample corresponding 
to an integrated luminosity of $208.7 \invfb$ collected at a center-of-mass 
energy $\sqrt{s}$ of 10.58 $\gev$ and $21.5 \invfb$ at $\sqrt{s} = 10.54 \gev$. 
With a cross-section for \tautau\ pair production averaged over $\sqrt{s}$ of 
$\sigma_{\scriptscriptstyle \mtau \mtau} = (0.919 \pm 0.003) \nb$~\cite{BanerjeeLumi},
this total data sample of $230.2 \invfb$ contains 211.6 million $\tau^+ \tau^-$ pairs.

Studies of \mc\ (MC) simulated events are carried out for signal and various background samples;
$\tau$ pairs are generated with \kk~\cite{Ward:2002qq} and their decays simulated with 
\tauola~\cite{Jadach:1993hs}. Signal $\tau$ decays are modeled using form-factors from the Breit-Wigner 
line-shape of \Kstareightninetwo\ decays~\cite{pdg:2006}, which nearly
saturates the $\tau^- \rightarrow K^- \pi^0 \nu_{\tau}$ final 
state~\cite{Battle:1994by,Barate:1999hi,Abbiendi:2004xa}.
The other $\tau$ decays into any of the possible final states, 
according to the measured branching fractions~\cite{pdg:2006}. 
To estimate non-tau backgrounds, samples of $\FourS \rightarrow \BB$, 
$e^+ e^- \rightarrow \qqbar$ ($q=u,d,s,c$) and $e^+ e^- \to \mu^+ \mu^-~(\gamma)$ are generated 
with \evtgen~\cite{Lange:2001uf}, \jetset~\cite{Sjostrand:1995iq} and \kk~\cite{Ward:2002qq} 
MC programs, respectively. The available MC samples are weighted according to their respective
size and cross-sections in order to match the data integrated luminosity~\cite{pdg:2006}.

Each event is divided into hemispheres in the center-of-mass (CM) 
frame using the plane perpendicular to the thrust axis, which is the direction that maximizes 
the sum of the longitudinal components of the momenta of reconstructed particles, both neutral 
and charged. Only events with one charged track in each hemisphere, and with both tracks consistent 
with originating from the interaction point (1-1 topology) are selected. The net charge of the event 
is required to be zero.

To suppress light quark ($u,d,s$) hadronic events, while retaining the relatively large fraction 
($\approx 35\%$) of $\tau$ leptons that decay leptonically, we require that one hemisphere 
contains a track that is identified either as an electron (\etag) or as a muon (\mutag).
The charged track in the opposite hemisphere is then required to be within the geometrical acceptance 
of the DIRC and to be identified as a kaon, and inconsistent with an electron. 
In order to reject events where the kaon has decayed or interacted before reaching the DIRC, a 
two-dimensional requirement on the \Cerenkov\ angle ($\theta_{C}$) versus the laboratory momentum 
of the candidate kaon ($p_{lab}$) is used: 
$\theta_{C}~(rad) \leq 0.48 +0.31*p_{lab}~(\gevc)$.

Event shape variables are used to discriminate against remaining \BB\ and \qqbar 
backgrounds. The thrust magnitude is required to be greater than $0.9$, and the ratio of the $2^{\rm nd}$ 
to the $0^{\rm th}$ Fox-Wolfram moments~\cite{Fox-Wolfram} is required to be greater than $0.5$. 
Also, to account for the substantial energy carried away by neutrinos in \mtau\-pair events, the total 
missing momentum in the laboratory frame is required to be greater than $0.5 \gevc$. This discriminates 
against Bhabha scattering and $\mu$-pair events,  as well as \qqbar\ production.
Moreover, we remove events in which a \KS\ decay to two charged pions is identified.

We further require that the event contains only one \piz.
Only $\pi^0$ mesons that have been reconstructed from two separate EMC clusters with an energy 
above $50\mev$, not associated with charged tracks, are considered in the analysis. Candidate $\pi^0$ mesons
are required to have an invariant mass in the range $100 < m_{\gamma \gamma} < 160\mevcc$ and an energy above 
$200\mev$ in the CM frame. 
The mass of the selected photon pair is constrained to the nominal $\pi^0$ mass~\cite{pdg:2006}
to improve the purity of the selected $\pi^0s$. 
The angle between the charged kaon and \piz\ momenta in the CM frame is required to be less than 
$1.0$ radian. 

After all the above requirements, there remain $44348$ $e$-tagged events, and $33764$ 
$\mu$-tagged events.

The MC simulated events are adjusted to improve their accuracy in modeling data events, 
according to several dedicated studies on specific control samples.
Charged tracks are weighted to compensate for the different particle identification (PID) efficiencies
between data and MC.
On average, the MC efficiency is reduced by 15\% and 3\% for muons and kaons, respectively. 
The electron identification is properly simulated and therefore no MC efficiency correction is applied.
A correction to the $\piz$ MC efficiency has been obtained from detailed studies based
on $\taum \to \rhom \nut$ and $\taum \to \pim \nut$ events. As a result of this study, the 
\piz\ MC efficiency is reduced by 2\%.

We estimate the \taumtoKmpiz\ selection efficiencies to be $1.31\%$, $0.96\%$ and 
$2.27\%$ for the \etag, \mutag\, and combined samples, respectively, using the signal MC sample
with all requirements and corrections applied. 
The efficiency as a function of the $K^{-} \pi^0$ mass is consistent with being constant.

\begin{figure}[tb]
\includegraphics[width=20pc,height=16pc]{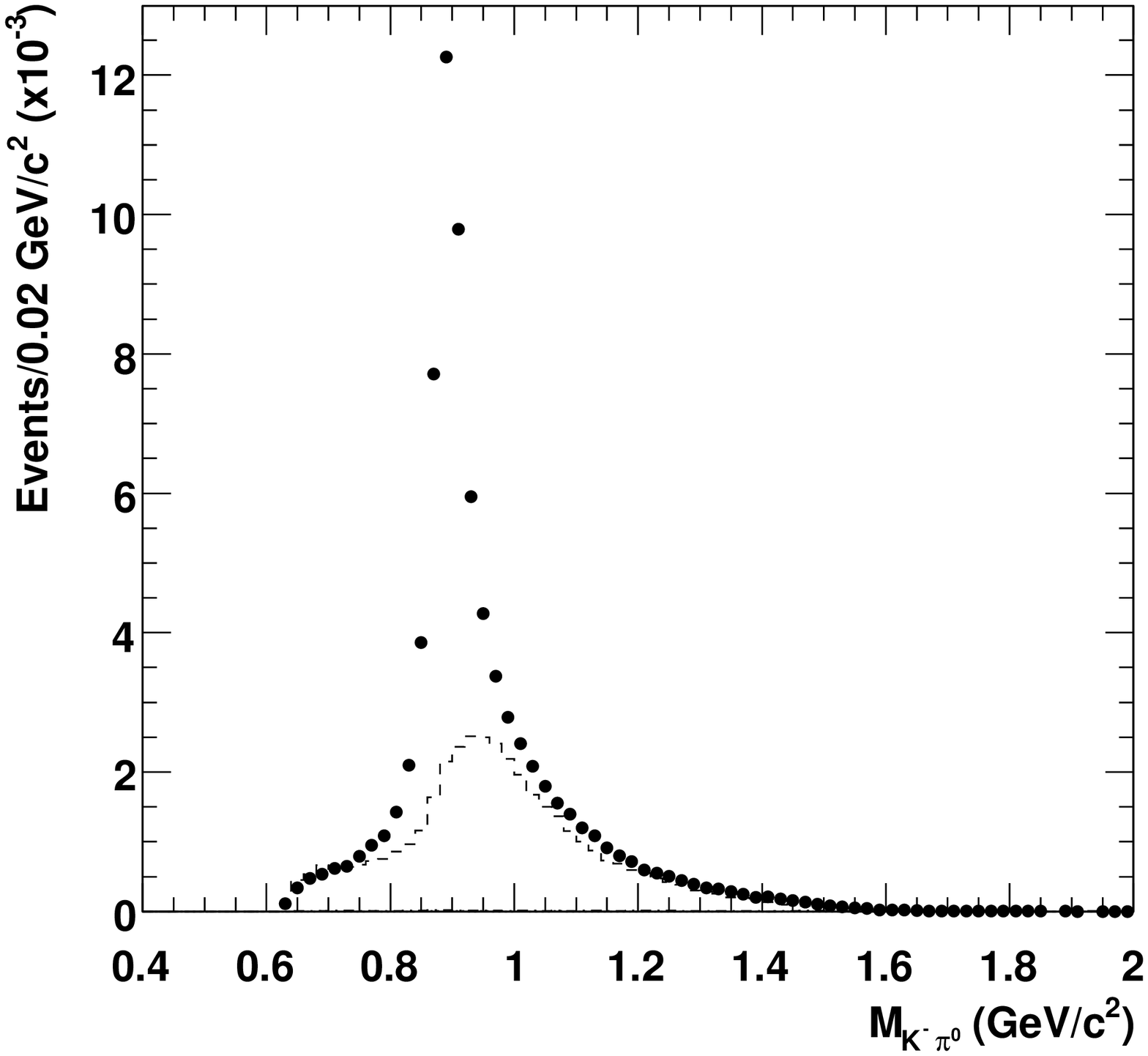} \\
\includegraphics[width=20pc,height=16pc]{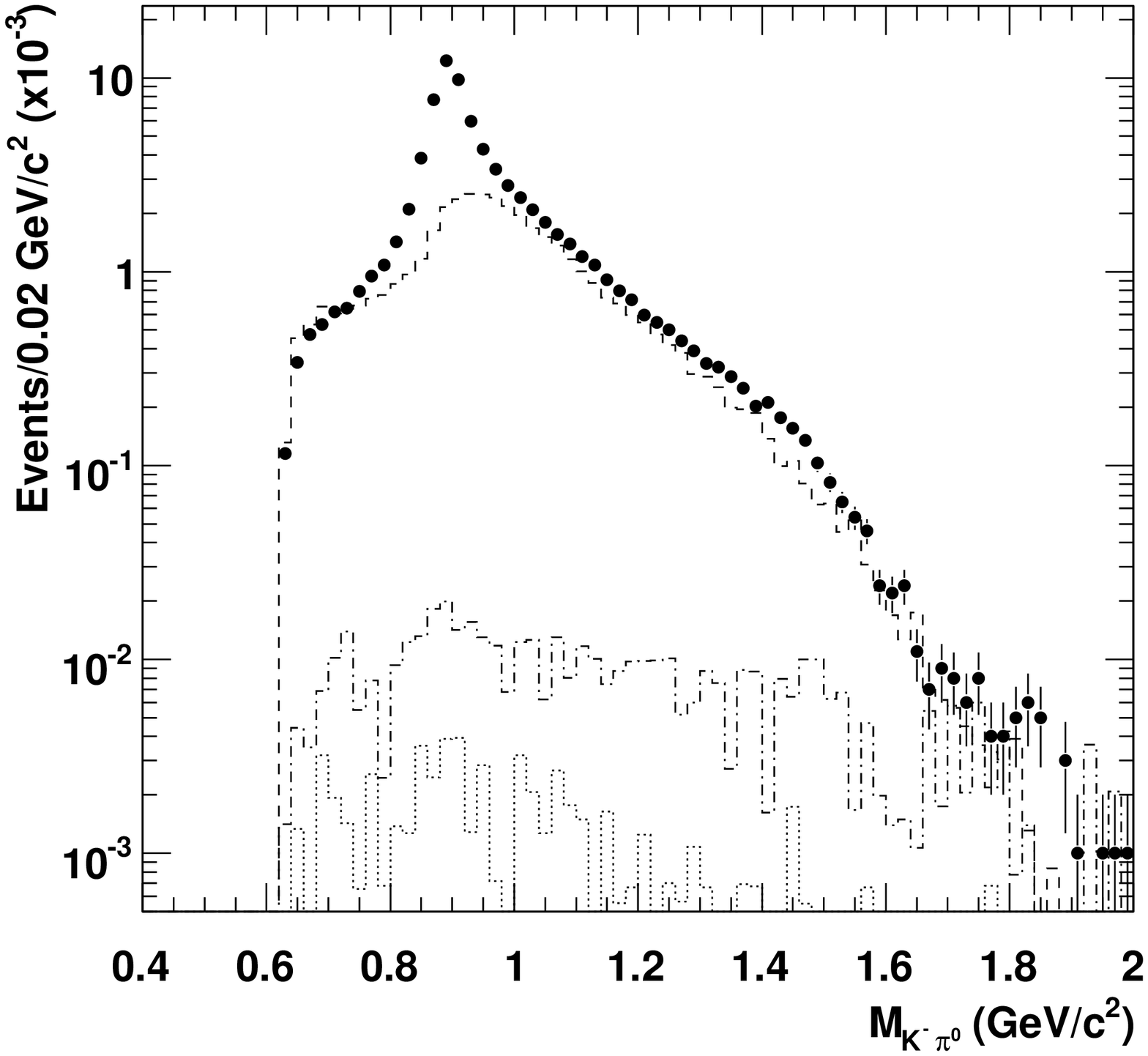} 
\caption{Distribution of the reconstructed $K^- \pi^0$ mass for the combined sample in linear (top) and logarithmic 
         (bottom) scale. 
         The dots are the data, while the histograms are background MC events with selection and efficiency corrections:
	 $\tau$ background (dashed line), $q\bar{q}$ (dash-dotted line), $\mu^+ \mu^-$ (dotted line).}
\label{fig_kpi0}
\end{figure}

Figure~\ref{fig_kpi0} shows the invariant mass spectra of the selected $K^- \pi^0$ candidates 
and simulated backgrounds for the combined sample after all the analysis requirements.
The contribution from light quarks, $\ccbar$ and $\mu^+ \mu^- (\gamma)$ backgrounds is much smaller than the $\tau$-pair
backgrounds. The contribution from \BB backgrounds is negligible. 
The $K^{*}(892)^-$ resonance is seen prominently above the simulated background.
Decays to higher $\Kstar$ resonances are expected near $1.4 \gevcc$
\cite{Finkemeier:1996dh,Maltman:2002wb,Gamiz:2004prl}, such as the $\Kstarfourteentenm$~\cite{Barate:1999hj} 
and $\Kstarfourteenthirtym$~\cite{Barate:1999hj}, but their branching fractions are not well measured yet. 
These decays are not included in our simulation of $\tau$-pair events, but appear to be present in the 
data near $1.4 \gevcc$. Below $0.7\gevcc$, the background is dominated by $K^- \pi^0 \pi^0$ and $K^- K^0 \pi^0$ events, 
for which the branching fractions are only known with relative uncertainties $\approx 40\%$ and $\approx 15\%$, 
respectively. These uncertainties are taken into account in the estimation of the systematic uncertainty due to 
modeling of the $\tau$ backgrounds. 

The branching fraction $\BRtaumtoKmpiz$ is calculated as
$\BR = [N_{\scriptscriptstyle{\rm sel}} - N_{\scriptscriptstyle{\rm bkg}}] / 2 \SigEff \Ntautau$,
where $\Ntautau = \lum ~\sigma_{\scriptscriptstyle \mtau \mtau} = 2.116 \times 10^8$ is 
the number of produced \tautau\ pairs; $\SigEff$ is the estimated signal selection efficiency;
$N_{\scriptscriptstyle{\rm sel}}$ is the number of events selected in data; and
$N_{\scriptscriptstyle{\rm bkg}}$ is the estimated number of background events obtained from MC 
simulations. For the combined $e$-tagged and $\mu$-tagged sample, 
$N_{\scriptscriptstyle{\rm sel}} = 78112 \pm 280$ events and the estimated background is 
$N_{\scriptscriptstyle{\rm bkg}} = 38247 \pm 159$ events. 
The branching fraction $\BRtaumtoKmpiz$ is found to be $(0.416 \pm 0.003 \, \stat)\%$, 
where the statistical uncertainty comes from the uncertainty in the number of signal events,
$N_{\scriptscriptstyle{\rm sel}} - N_{\scriptscriptstyle{\rm bkg}}$.
Several cross checks were performed by determining the branching fraction as a function of tag type, 
$\pi^0$ momentum, kaon charge and run period; all were found to give consistent results.

A summary of the systematic uncertainties is given in Table~\ref{tab:kpi0systematics}.
The uncertainty in the charged track reconstruction efficiency is estimated to be $0.31\%$ per track,
based on studies of data control samples of $\tau$-pair events decaying to one charged particle
on one side and three charged particles on the other side. 
The systematic uncertainty associated with the efficiency of detecting a $\pi^0$ is $3.26\%$. 
The stated uncertainty in the charged particle identification efficiency represents the combined
uncertainty for the two charged tracks, ($e$, $K$) or ($\mu$, $K$). 
This uncertainty includes a contribution due to the mis-identification of charged pions as kaons.
The uncertainty associated with the \tautau\-pair production \crosssection\ is 
$0.31\%$~\cite{BanerjeeLumi} and the luminosity determination uncertainty is $0.94\%$. 
The effects of approximations in the MC signal modeling and of the finite MC statistics 
on the overall efficiency are negligible, but have been included in the estimation
of the systematic uncertainty.
The branching fractions for several \mtau decay modes that contribute to the background, particularly 
Cabibbo-suppressed decays, are not well known.
The resulting uncertainty due to the $\tau$-pair background estimate
on the $\tau^- \rightarrow K^- \pi^0 \nu_{\tau}$ branching fraction is 1.35\%.
Backgrounds from other sources are very small and their impact of the signal extraction is negligible.

The total systematic uncertainty is the quadratic sum of the individual sources described above, 
and is 4.32\% for the combined sample.

\begin{table}[tb]
\caption{\small Relative systematic uncertainties in the $\taum \to \Km \piz \nut$ analysis.}
\label{tab:kpi0systematics}
\begin{small}
\begin{center}
\begin{tabular}{c||c|c|c} \hline
Systematic                                       & $e$-tag& $\mu$-tag & Combined \\
                                                 &   (\%) &   (\%) & (\%) \\ \hline \hline
Tracking efficiency                              & $0.62$ & $0.62$ & $0.62$ \\
$\piz$ efficiency                                & $3.26$ & $3.26$ & $3.26$ \\ 
Particle identification                          & $2.09$ & $2.34$ & $2.17$ \\ 
$\tau \tau$ cross section ($\sigma_{\tau \tau}$) & $0.31$ & $0.31$ & $0.31$ \\
Luminosity ($\lum$)                              & $0.94$ & $0.94$ & $0.94$ \\
Signal modeling \& MC statistics                 & $0.38$ & $0.52$ & $0.35$ \\ 
\taubkgds                                        & $1.35$ & $1.35$ & $1.35$ \\ \hline
TOTAL                                            & $4.28$ & $4.42$ & $4.32$ \\ \hline
\end{tabular}
\end{center} 
\end{small}
\end{table}

In summary, using 211.6 million $\tau^+~\tau^-$ pairs recorded by the \babar\ detector,
we obtain the following result:
\begin{eqnarray}
\nonumber
\lefteqn{\BRtaumtoKmpiz =} 
\\ && \mbox{}
\nonumber
\left( 0.416 \pm 0.003 \, \stat \pm 0.018 \, \syst \right) \%.
\end{eqnarray}
This measurement of the branching fraction is the most precise to date
and is consistent with the existing world average,
$\BRtaumtoKmpiz = (0.454 \pm 0.030)\%$~\cite{pdg:2006}.

\section{Acknowledgements}

We are grateful for the 
extraordinary contributions of our \pep2\ colleagues in
achieving the excellent luminosity and machine conditions
that have made this work possible.
The success of this project also relies critically on the 
expertise and dedication of the computing organizations that 
support \babar.
The collaborating institutions wish to thank 
SLAC for its support and the kind hospitality extended to them. 
This work is supported by the
US Department of Energy
and National Science Foundation, the
Natural Sciences and Engineering Research Council (Canada),
the Commissariat \`a l'Energie Atomique and
Institut National de Physique Nucl\'eaire et de Physique des Particules
(France), the
Bundesministerium f\"ur Bildung und Forschung and
Deutsche Forschungsgemeinschaft
(Germany), the
Istituto Nazionale di Fisica Nucleare (Italy),
the Foundation for Fundamental Research on Matter (The Netherlands),
the Research Council of Norway, the
Ministry of Science and Technology of the Russian Federation, 
Ministerio de Educaci\'on y Ciencia (Spain), and the
Science and Technology Facilities Council (United Kingdom).
Individuals have received support from 
the Marie-Curie IEF program (European Union) and
the A. P. Sloan Foundation.


\begin{thebibliography}{21}
\expandafter\ifx\csname natexlab\endcsname\relax\def\natexlab#1{#1}\fi
\expandafter\ifx\csname bibnamefont\endcsname\relax
  \def\bibnamefont#1{#1}\fi
\expandafter\ifx\csname bibfnamefont\endcsname\relax
  \def\bibfnamefont#1{#1}\fi
\expandafter\ifx\csname citenamefont\endcsname\relax
  \def\citenamefont#1{#1}\fi
\expandafter\ifx\csname url\endcsname\relax
  \def\url#1{\texttt{#1}}\fi
\expandafter\ifx\csname urlprefix\endcsname\relax\def\urlprefix{URL }\fi
\providecommand{\bibinfo}[2]{#2}
\providecommand{\eprint}[2][]{\url{#2}}

\bibitem[{\citenamefont{Cabibbo}(1963)}]{Cabibbo:1963yz}
\bibinfo{author}{\bibfnamefont{N.}~\bibnamefont{Cabibbo}},
  \bibinfo{journal}{Phys. Rev. Lett.} \textbf{\bibinfo{volume}{10}},
  \bibinfo{pages}{531} (\bibinfo{year}{1963}).

\bibitem[{\citenamefont{Kobayashi and Maskawa}(1973)}]{Kobayashi:1973fv}
\bibinfo{author}{\bibfnamefont{M.}~\bibnamefont{Kobayashi}} \bibnamefont{and}
  \bibinfo{author}{\bibfnamefont{T.}~\bibnamefont{Maskawa}},
  \bibinfo{journal}{Prog. Theor. Phys.} \textbf{\bibinfo{volume}{49}},
  \bibinfo{pages}{652} (\bibinfo{year}{1973}).

\bibitem[{\citenamefont{Jamin et~al.}(2006)\citenamefont{Jamin, Oller, and
  Pich}}]{Jamin:2006tj}
\bibinfo{author}{\bibfnamefont{M.}~\bibnamefont{Jamin}},
  \bibinfo{author}{\bibfnamefont{J.~A.} \bibnamefont{Oller}}, \bibnamefont{and}
  \bibinfo{author}{\bibfnamefont{A.}~\bibnamefont{Pich}},
  \bibinfo{journal}{Phys. Rev. D} \textbf{\bibinfo{volume}{74}},
  \bibinfo{pages}{074009} (\bibinfo{year}{2006}).

\bibitem[{\citenamefont{Gamiz et~al.}(2003)\citenamefont{Gamiz, Jamin, Pich,
  Prades, and Schwab}}]{Gamiz:2002nu}
\bibinfo{author}{\bibfnamefont{E.}~\bibnamefont{Gamiz}},
  \bibinfo{author}{\bibfnamefont{M.}~\bibnamefont{Jamin}},
  \bibinfo{author}{\bibfnamefont{A.}~\bibnamefont{Pich}},
  \bibinfo{author}{\bibfnamefont{J.}~\bibnamefont{Prades}}, \bibnamefont{and}
  \bibinfo{author}{\bibfnamefont{F.}~\bibnamefont{Schwab}},
  \bibinfo{journal}{JHEP} \textbf{\bibinfo{volume}{01}}, \bibinfo{pages}{060}
  (\bibinfo{year}{2003}).

\bibitem[{\citenamefont{Maltman and Wolfe}(2007)}]{Maltman:2007pr}
\bibinfo{author}{\bibfnamefont{K.}~\bibnamefont{Maltman}} \bibnamefont{and}
  \bibinfo{author}{\bibfnamefont{C.~E.} \bibnamefont{Wolfe}}
  (\bibinfo{year}{2007}), \eprint{arXiv:hep-ph/0703314}.

\bibitem[{\citenamefont{Battle et~al.}(1994)}]{Battle:1994by}
\bibinfo{author}{\bibfnamefont{M.}~\bibnamefont{Battle}} \bibnamefont{et~al.}
  (\bibinfo{collaboration}{CLEO}), \bibinfo{journal}{Phys. Rev. Lett.}
  \textbf{\bibinfo{volume}{73}}, \bibinfo{pages}{1079} (\bibinfo{year}{1994}).

\bibitem[{\citenamefont{Barate et~al.}(1999{\natexlab{a}})}]{Barate:1999hi}
\bibinfo{author}{\bibfnamefont{R.}~\bibnamefont{Barate}} \bibnamefont{et~al.}
  (\bibinfo{collaboration}{ALEPH}), \bibinfo{journal}{Eur. Phys. J. C}
  \textbf{\bibinfo{volume}{10}}, \bibinfo{pages}{1}
  (\bibinfo{year}{1999}{\natexlab{a}}).

\bibitem[{\citenamefont{Abbiendi et~al.}(2004)}]{Abbiendi:2004xa}
\bibinfo{author}{\bibfnamefont{G.}~\bibnamefont{Abbiendi}} \bibnamefont{et~al.}
  (\bibinfo{collaboration}{OPAL}), \bibinfo{journal}{Eur. Phys. J. C}
  \textbf{\bibinfo{volume}{35}}, \bibinfo{pages}{437} (\bibinfo{year}{2004}).

\bibitem[{\citenamefont{{\pep2\ collaboration}}(1993)}]{:1993mp}
\bibinfo{author}{\bibnamefont{{\pep2\ collaboration}}} (\bibinfo{year}{1993}),
  \eprint{SLAC-418}.

\bibitem[{\citenamefont{Aubert et~al.}(2002)}]{Aubert:2001tu}
\bibinfo{author}{\bibfnamefont{B.}~\bibnamefont{Aubert}} \bibnamefont{et~al.}
  (\bibinfo{collaboration}{BABAR}), \bibinfo{journal}{Nucl. Instrum. Meth. A}
  \textbf{\bibinfo{volume}{479}}, \bibinfo{pages}{1} (\bibinfo{year}{2002}).

\bibitem[{\citenamefont{Banerjee et~al.}()\citenamefont{Banerjee, Pietrzyk,
  Roney, and Was}}]{BanerjeeLumi}
\bibinfo{author}{\bibfnamefont{S.}~\bibnamefont{Banerjee}},
  \bibinfo{author}{\bibfnamefont{B.}~\bibnamefont{Pietrzyk}},
  \bibinfo{author}{\bibfnamefont{J.~M.} \bibnamefont{Roney}}, \bibnamefont{and}
  \bibinfo{author}{\bibfnamefont{Z.}~\bibnamefont{Was}},
  \eprint{arXiv:0706.3235 [hep-ph]}.

\bibitem[{\citenamefont{Ward et~al.}(2003)\citenamefont{Ward, Jadach, and
  Was}}]{Ward:2002qq}
\bibinfo{author}{\bibfnamefont{B.~F.~L.} \bibnamefont{Ward}},
  \bibinfo{author}{\bibfnamefont{S.}~\bibnamefont{Jadach}}, \bibnamefont{and}
  \bibinfo{author}{\bibfnamefont{Z.}~\bibnamefont{Was}},
  \bibinfo{journal}{Nucl. Phys. Proc. Suppl.} \textbf{\bibinfo{volume}{116}},
  \bibinfo{pages}{73} (\bibinfo{year}{2003}).

\bibitem[{\citenamefont{Jadach et~al.}(1993)\citenamefont{Jadach, Was, Decker,
  and K{\"u}hn}}]{Jadach:1993hs}
\bibinfo{author}{\bibfnamefont{S.}~\bibnamefont{Jadach}},
  \bibinfo{author}{\bibfnamefont{Z.}~\bibnamefont{Was}},
  \bibinfo{author}{\bibfnamefont{R.}~\bibnamefont{Decker}}, \bibnamefont{and}
  \bibinfo{author}{\bibfnamefont{J.~H.} \bibnamefont{K{\"u}hn}},
  \bibinfo{journal}{Comput. Phys. Commun.} \textbf{\bibinfo{volume}{76}},
  \bibinfo{pages}{361} (\bibinfo{year}{1993}).

\bibitem[{\citenamefont{Yao. et~al.}(2006)}]{pdg:2006}
\bibinfo{author}{\bibfnamefont{W.-M.} \bibnamefont{Yao.}} \bibnamefont{et~al.}
  (\bibinfo{collaboration}{Particle Data Group}), \bibinfo{journal}{Journal of
  Physics G} \textbf{\bibinfo{volume}{33}}, \bibinfo{pages}{1}
  (\bibinfo{year}{2006}).

\bibitem[{\citenamefont{Lange}(2001)}]{Lange:2001uf}
\bibinfo{author}{\bibfnamefont{D.~J.} \bibnamefont{Lange}},
  \bibinfo{journal}{Nucl. Instrum. Meth. A} \textbf{\bibinfo{volume}{462}},
  \bibinfo{pages}{152} (\bibinfo{year}{2001}).

\bibitem[{\citenamefont{Sj{\"o}strand}(1994)}]{Sjostrand:1995iq}
\bibinfo{author}{\bibfnamefont{T.}~\bibnamefont{Sj{\"o}strand}},
  \bibinfo{journal}{Comput. Phys. Commun.} \textbf{\bibinfo{volume}{82}},
  \bibinfo{pages}{74} (\bibinfo{year}{1994}).

\bibitem[{\citenamefont{Fox and Wolfram}(1979)}]{Fox-Wolfram}
\bibinfo{author}{\bibfnamefont{G.~C.} \bibnamefont{Fox}} \bibnamefont{and}
  \bibinfo{author}{\bibfnamefont{S.}~\bibnamefont{Wolfram}},
  \bibinfo{journal}{Nucl. Phys.} \textbf{\bibinfo{volume}{B149}},
  \bibinfo{pages}{413} (\bibinfo{year}{1979}).

\bibitem[{\citenamefont{Finkemeier and Mirkes}(1996)}]{Finkemeier:1996dh}
\bibinfo{author}{\bibfnamefont{M.}~\bibnamefont{Finkemeier}} \bibnamefont{and}
  \bibinfo{author}{\bibfnamefont{E.}~\bibnamefont{Mirkes}},
  \bibinfo{journal}{Z. Phys. C} \textbf{\bibinfo{volume}{72}},
  \bibinfo{pages}{619} (\bibinfo{year}{1996}).

\bibitem[{\citenamefont{Maltman}(2003)}]{Maltman:2002wb}
\bibinfo{author}{\bibfnamefont{K.}~\bibnamefont{Maltman}},
  \bibinfo{journal}{Nucl. Phys. Proc. Suppl.} \textbf{\bibinfo{volume}{123}},
  \bibinfo{pages}{149} (\bibinfo{year}{2003}).

\bibitem[{\citenamefont{Gamiz et~al.}(2005)\citenamefont{Gamiz, Jamin, Pich,
  Prades, and Schwab}}]{Gamiz:2004prl}
\bibinfo{author}{\bibfnamefont{E.}~\bibnamefont{Gamiz}},
  \bibinfo{author}{\bibfnamefont{M.}~\bibnamefont{Jamin}},
  \bibinfo{author}{\bibfnamefont{A.}~\bibnamefont{Pich}},
  \bibinfo{author}{\bibfnamefont{J.}~\bibnamefont{Prades}}, \bibnamefont{and}
  \bibinfo{author}{\bibfnamefont{F.}~\bibnamefont{Schwab}},
  \bibinfo{journal}{Phys. Rev. Lett.} \textbf{\bibinfo{volume}{94}},
  \bibinfo{pages}{011803} (\bibinfo{year}{2005}).

\bibitem[{\citenamefont{Barate et~al.}(1999{\natexlab{b}})}]{Barate:1999hj}
\bibinfo{author}{\bibfnamefont{R.}~\bibnamefont{Barate}} \bibnamefont{et~al.}
  (\bibinfo{collaboration}{ALEPH}), \bibinfo{journal}{Eur. Phys. J. C}
  \textbf{\bibinfo{volume}{11}}, \bibinfo{pages}{599}
  (\bibinfo{year}{1999}{\natexlab{b}}).

\end{thebibliography}
\end{document}